\documentclass{article} 
\usepackage{nips11submit_e,times}

\usepackage{amsmath}
\usepackage{amsthm}
\usepackage{graphicx}
\usepackage{subfig}
\usepackage{url}

\usepackage[compact]{titlesec}

\usepackage{xspace}
\newcommand{\pmone}{\{-1,1\}^n\xspace}
\newcommand{\zrone}{\{0,1\}^n\xspace}
\newcommand{\ie}{\emph{i.e.},\xspace}

\newtheorem{theorem}{Theorem}

\newtheorem{conjecture}[theorem]{Conjecture}

\usepackage{comment}

\title{Recovery of a Sparse Integer Solution to an Underdetermined System
of Linear Equations}

\author{
T.~S.~Jayram \\
IBM Research - Almaden\\
\texttt{\small jayram@almaden.ibm.com}
\And
Soumitra~Pal \\
CSE, IIT - Bombay \\
\texttt{\small mitra@cse.iitb.ac.in}
\And
Vijay~Arya \\
IBM Research - India \\
\texttt{\small vijay.arya@in.ibm.com}
}

\nipsfinalcopy 

\begin{document}


\maketitle


\begin{abstract}
We consider a system of $m$ linear equations in $n$ variables $Ax=b$ where $A$ is a given $m \times n$ matrix and 
$b$ is a given $m$-vector known to be equal to $A\bar{x}$ for some unknown solution $\bar{x}$ that is \emph
{integer} and \emph{$k$-sparse}: $\bar{x} \in \zrone$ and exactly $k$ entries of $\bar{x}$ are~1. We give 
necessary and sufficient conditions for recovering the solution $\bar{x}$ exactly using an LP relaxation that 
minimizes $\ell_1$ norm of $x$. When $A$ is drawn from a distribution that has exchangeable columns, 
we show an interesting connection
between the recovery probability and a well known problem in geometry, namely the $k$-set problem. 
To the best of our knowledge, this connection appears to be new in the compressive sensing literature. We 
empirically show that for large $n$ if the elements of $A$ are drawn i.i.d. from the normal distribution then the 
performance of the recovery LP exhibits a \emph{phase transition}, 
\ie for each $k$ there exists a value $\bar{m}$ 
of $m$ such that the recovery always succeeds if $m > \bar{m}$ and always fails if $m < \bar{m}$. Using the 
empirical data we conjecture that $\bar{m} = nH(k/n)/2$ where $H(x) = -x \log_2x - (1-x)\log_2(1-x)$ is the 
binary entropy function.

\end{abstract}

\section{Introduction}

We consider the system of linear equations in the real vector variable $x$:
\begin{align}
  Ax = b \label{main}
\end{align}
where $A$ is a given real $m \times n$ matrix, $b$ is a given vector in $R^m$ and $x \in R^n$.  Suppose it is known that the system has an underlying solution that is \emph{binary}, \ie $\exists ~\bar{x} \in \zrone$ such that $b = A\bar{x}$.  We are interested in the conditions under which $\bar{x}$ can be recovered exactly and efficiently. 

The above problem occurs in smart grids where we wish to retrieve the underlying phase connectivity from a time series of meter measurements~\cite{vijay2010phase}.  Each customer household is connected to one of the three phases of a low-voltage transformer that distributes power to households.  Both households and transformers have smart meters. Therefore for a series of time intervals, it is known as to how many watt-hours of power are sent out on each phase and how many are consumed by each customer.  However, to which phase a customer is connected is not known. System~\eqref{main} is the problem formulation for a single phase based on the principles of conservation of power.  The columns of $A$ hold the time series of meter measurements from customers, $b$ holds the corresponding time series for a phase, and $x$ determines if a customer is connected to that phase. Empirical data collected from real smart meters shows that measurements have sufficient variability over time and customers implying that $A$ has full rank.

If $m=n$, the unique underlying binary solution to~\eqref{main} is recovered as $\bar{x} = A^{-1}b$.  If $m<n$, system~\eqref{main} has infinite real solutions and may have multiple binary solutions.  When $m=1$, the problem reduces to Subset-Sum problem, which is NP-hard.  For $m < n$, even if a binary solution to~\eqref{main} is given, checking if it is a unique solution is also NP-hard~\cite{papadimitriou1984complexity}.


To circumvent these difficulties, Mangasarian et al.~\cite{mangasarian2011probability} first transform~\eqref{main} to its equivalent $Ay = d$ using $y = e - 2x$ where $d = Ae - 2b$ and $e$ is a column vector of all ones. Then they give necessary and sufficient conditions for the uniqueness of an integer solution $y \in \pmone$ to the following LP relaxation that minimizes the $\ell_{\infty}$ norm of $y$:
\begin{align}
  \text{min} \; \delta \quad \text{s.t.} \quad Ay = d, \quad -\delta e \le y \le \delta e. \label{mlp}
\end{align}
A solution $\bar{y}$ that is unique to LP~\eqref{mlp} and is integer guarantees that $(e - \bar{y})/2$ exactly recovers $\bar{x}$. 
The paper~\cite{mangasarian2011probability} also computes the probability that a randomly generated problem instance of LP~\eqref{mlp} satisfies the uniqueness conditions. This gives a lower bound on the probability that~\eqref{main} has a unique binary solution. For large $n$, a transition behavior is observed: the probability of uniqueness is almost $0$  for $m/n < 1/2$ and almost 1 for $m/n > 1/2$. 



In this work, we follow the approach of~\cite{mangasarian2011probability} and study the conditions under which a \emph{binary} and \emph{$k$-sparse} $\bar{x}$ with exactly $k$ non-zero entries is a unique solution to the following alternate LP relaxation that minimizes the $\ell_1$ norm of $x$:
\begin{align}
  \text{min} \; e^T x \quad \text{s.t.} \quad Ax = b, \quad 0 \le x \le 1 \label{lp}
\end{align}
so that  \eqref{lp} exactly recovers the solution $\bar{x}$ of ~\eqref{main}. As in~\cite{mangasarian2011probability}, we wish to find necessary and sufficient conditions for the uniqueness of a binary $k$-sparse solution and to compute the probability of the conditions getting satisfied on a random instance as a function of $n, m, k$.  

Donoho et al.~\cite{donoho2010precise} also consider recovery of sparse solutions to~\eqref{main}, albeit with a different notion of sparsity. They consider three LP relaxations of which two are relevant to our work:
\begin{align}
  \text{min} \; 0^T x \quad \text{s.t.} \quad Ax &= b, \quad 0 \le x \le 1
  \quad \text{and}  \label{dlp2}\\
  \text{min} \; e^T x \quad \text{s.t.} \quad Ax &= b, \quad x \ge 0. 
 \label{dlp1} 
\end{align}
Donoho et al.'s  definition of sparsity is closely tied to the polytope defined by the constraints of the LP relaxation. 
A signal is considered $k$-sparse if it lies on a $k$-face of the constraint polytope. 
However in LP~\eqref{lp} a $k$-sparse binary signal does not lie on the $k$-face of its associated polytope. 
Therefore their techniques of counting faces of polytopes to compute uniqueness probabilities give us partial results, however they are not tight. 

For example, in case of LP~\eqref{dlp2}, $x$ is considered $k$-sparse if it has $n-k$ entries either $0$ or $1$ and $k$ entries in $(0, 1)$. Any binary signal lies on a vertex of LP~\eqref{dlp2}'s constraints polytope $0 \le x \le 1$ and has zero sparsity. Donoho et al. show that LP~\eqref{dlp2} requires $m/n = 1/2$ to recover a signal of zero sparsity (see figure 3 of ~\cite{donoho2010precise}, $Q=I$). This is not the strongest result possible because LP (\ref{lp}) can recover certain binary signals with $m/n$ strictly less than $1/2$. 
In case of LP~\eqref{dlp1}, $x$ is considered $k$-sparse if $n-k$ entries are $0$ and $k$ entries $>0$. In this case although their definition of sparsity coincides with ours, the results for recovery are not tight. For $m/n = 1/2$, LP~\eqref{dlp1} recovers a binary signal of sparsity at most $k=n/4$ (see figure 3 of~\cite{donoho2010precise}, $Q=T$) while LP~\eqref{lp} can even recover a binary signal of sparsity $k=n/2$.

The rest of the paper is organized as follows. Section~\ref{sec:conditions} gives necessary and sufficient conditions for a binary and $k$-sparse signal to be the unique solution of LP~\eqref{lp}. 
Section~\ref{sec:probability} links the probability of uniqueness  with the expected number of $k$-sets in a random set of points from $R^m$.
The latter problem is still open. 
Section~\ref{sec:results} presents experimental results and compares the performance of LPs.

\section{Conditions for a $k$-sparse $\bar{x} \in \zrone$ to be the unique solution of LP~\eqref{lp}}
\label{sec:conditions}

We use the necessary and sufficient conditions for a given solution to be the unique optimal solution of an LP. These are derived in~\cite{mangasarian1979uniqueness} and shown below for reference. 

\begin{theorem}[\cite{mangasarian1979uniqueness}, Theorem 2(iii)]
  Let $\bar{x}$ be a solution of the LP:
  \begin{align}
    \text{min} \; c^T x \quad \text{s.t.} \quad Gx = h, \quad Px \ge q. \label{genlp}
  \end{align}
  Let $P_{eq}$ denote the submatrix of $P$ consisting of rows of $Px \ge q$ which are tight at $\bar{x}$, \ie all rows $i$ such that $p_i \bar{x} = q_i$. Then $\bar{x}$ is unique if and only if there exists no $z$ satisfying 
  \begin{equation}
    Gz = 0, \quad P_{eq}z \ge 0, \quad c^T z \le 0, \quad z \ne 0. \label{conlp}
  \end{equation}
  \label{thm:unique}
\end{theorem}
\vspace{-20pt}
Using the results of Theorem~\ref{thm:unique}, we establish our main theoretical result:

\begin{theorem}
  Let $\bar{x} \in \zrone$ be a $k$-sparse solution of \eqref{main}, $J_0 = \{j: \bar{x}_j=0\}$ and $J_1 = [n] \setminus J_0$. Consider the points in $R^m$ corresponding to the columns of $A$. Let the points corresponding to $Q_0=\{A_j: j \in J_0\}$ be colored red and the rest of the points $Q_1=\{A_j: j \in J_0\}$ be colored green. Then $\bar{x}$ is the unique solution to LP~\eqref{lp} if and only if there exists a hyperplane in $R^m$ not passing through the origin that strictly separates the red points $Q_0$ from the green points $Q_1$.
  \label{thm:conditions}
\end{theorem}

\vspace{-5mm}
\begin{proof}
  We rewrite LP~\eqref{lp} in the form of~\eqref{genlp}:
  \begin{align*}
    \min \quad e^Tx \quad \text{s.t.} \quad Ax = b, \quad x \ge 0, \quad -x \ge -e.
\end{align*}
  The inequalities that are tight at $\bar{x}$ are $x_j \ge 0$ for all $j \in J_0$ and $-x_j \ge -1$ for all $j \in J_1$. Thus by Theorem~\ref{thm:unique}, $\bar{x}$ is a unique solution if and only if there is no solution to the following system:
  \begin{align*}
    Az = 0, \quad z_j \ge 0 \; \forall j \in J_0, \quad -z_j \ge 0 \;
    \forall j \in J_1, \quad e^T z \le 0, \quad z \ne 0
  \end{align*}
  which is equivalent to the following system using substitution $y_j = z_j$ if $j \in J_0$, $-z_j$ otherwise: 
  \begin{align}
    \sum_{j \in J_0} y_j A_j = \sum_{j \in J_1} y_j A_j, \quad y_j \ge 0
    \; \forall j, \quad \sum_{j \in J_0} y_j \le \sum_{j \in J_1} y_j, \quad y \ne 0.
    \label{intermediate}
  \end{align}

  Introduce a new variable $y_0 \ge 0$ in $J_0$ with the column $A_0 = 0$. Then system~\eqref{intermediate} is equivalent to:
  \begin{align*}
    \sum_{j \in J_0} y_j A_j = \sum_{j \in J_1} y_j A_j, \quad
    \sum_{j \in J_0} y_j = \sum_{j \in J_1} y_j, \quad
    y \ge 0, \quad y \ne 0
  \end{align*}
  which is equivalent to the following system using the substitution $\alpha_j = y_j/\sum_{j\in J_t} y_j$ for $j \in J_t$:
  \begin{align*}
    \sum_{j \in J_0} \alpha_j A_j = \sum_{j \in J_1} \alpha_j A_j, \quad
    \sum_{j \in J_0} \alpha_j = \sum_{j \in J_1} \alpha_j = 1, \quad
    \alpha \ge 0, \quad \alpha \ne 0.
  \end{align*}

  Thus $\bar{x}$ is the unique optimal solution of LP~\eqref{lp} if and only if there exist no common points between the two convex sets $conv(Q_0 \cup \{0\})$ and $conv(Q_1)$, \ie there exists a hyperplane not passing through origin that strictly separates $Q_0$ from $Q_1$. 
\end{proof}

\section{Probability of uniqueness for LP~\eqref{lp}}
\label{sec:probability}

We obtain an initial expression for $P_{m,n,k}$, the probability that a hyperplane separates the two point sets $Q_0 \cup \{0\}$ and $Q_1$ given by an arbitrary but fixed $k$-sparse $\bar{x} \in \zrone$ for a random $A \in R^{m\times n}$. A \emph{$k$-set} of a finite point set $S$ in the euclidean space is a subset of $k$ elements of $S$ that can be strictly separated from the remaining points by a hyperplane.
\begin{theorem}
Suppose $A$ is drawn from a distribution that has exchangeable columns, i.e. every permutation $\pi$
applied to the columns of $A$ leaves the distribution of $A$ unchanged.
If $X$ is the random variable denoting the number of $k$-sets of the points in $R^m$ 
corresponding to the columns of $A$ then $P_{m,n,k}= E[X]/{n \choose k}$.
\end{theorem}

\vspace{-15pt}

\begin{proof}
  Let $t = {n \choose k}$, $S_1, S_2, \ldots, S_t$ be all possible subsets of columns of $A$ of size $k$ and $X_i$ be the indicator random variable if $S_i$ is indeed a $k$-set. Then $X = \sum_{i=1}^t X_i$ and $E[X] = \sum_{i=1}^t E[X_i] = \sum_{i=1}^t P[X_i=1]$. Since the distribution of the columns of $A$ remains unchanged after any  permutation, we can show that any property that is satisfied on a subset of size $k$ of the columns is also equally probable to be satisfied on another $k$-subset. Hence any subset of size $k$ is equally likely to be a $k$-set and hence has probability $P_{m,n,k}$. Thus $E[X] = \sum_{i=1}^t P_{m,n,k} = t P_{m,n,k}$ implying that $P_{m,n,k}= E[X]/{n \choose k}$.
\end{proof}

\vspace{-10pt}

Finding the number of $k$-sets of an arbitrary set of points is a long standing open problem~\cite{toth2001point}. The paper~\cite{clarksonexpected} is the only known work that gives an upper bound on the expected number of $k$-sets of a random set of points. However we seek a good lower bound. 

\section{Experimental results}
\label{sec:results}

We conduct Monte-Carlo simulations and test LPs~\eqref{mlp},~\eqref{lp} and~\eqref{dlp1} for recovery. A single simulation experiment with parameters $F, n, m, k,$ and $D$ consists of generating a random $m \times n$ matrix $A$ with entries from distribution $D$, generating two random solutions $\bar{x} \in \zrone, \bar{y} \in \pmone$ each having exactly $k$ $1$'s, computing $b=A\bar{x}, d=A\bar{y}$ and solving the LP $F$. We record a success if the solution of $F$ matches the solution we started with, with a relative error of $10^{-9}$. For each combination of $n, m, k, D$ we consider $200$ random instances of $(A, \bar{x}, \bar{y})$ and record the success rate as the fraction $R_F$ of instances for which LP~$F$ succeeds.  
We repeat the experiments by varying $n, m, k, D$. We consider the following distributions for $D$. $D_1$: each entry of $A$ comes from standard normal distribution $\mathcal{N}(\mu=0, \sigma^2{=}1)$, $D_2$: $\mathcal{N}(100, 1)$, $D_3$: Uniform distribution $\mathcal{U}(0,100)$, and $D_4$: each column $A_j$ of $A$ comes from $\mathcal{N}(\mu_j, 1)$ where $\mu_j \sim \mathcal{U}(0, 100)$. We experiment with $n=\{200, 500, 1600\}$ and for each $n$ we vary $m$ and $k$.

\begin{figure}[h!tb]
  \centering
  \subfloat[Transition behavior of LPs~\eqref{lp},\eqref{dlp1} in $(\rho,\delta)$ space]{
    \label{fig:left}
    \includegraphics[width=0.49\textwidth]{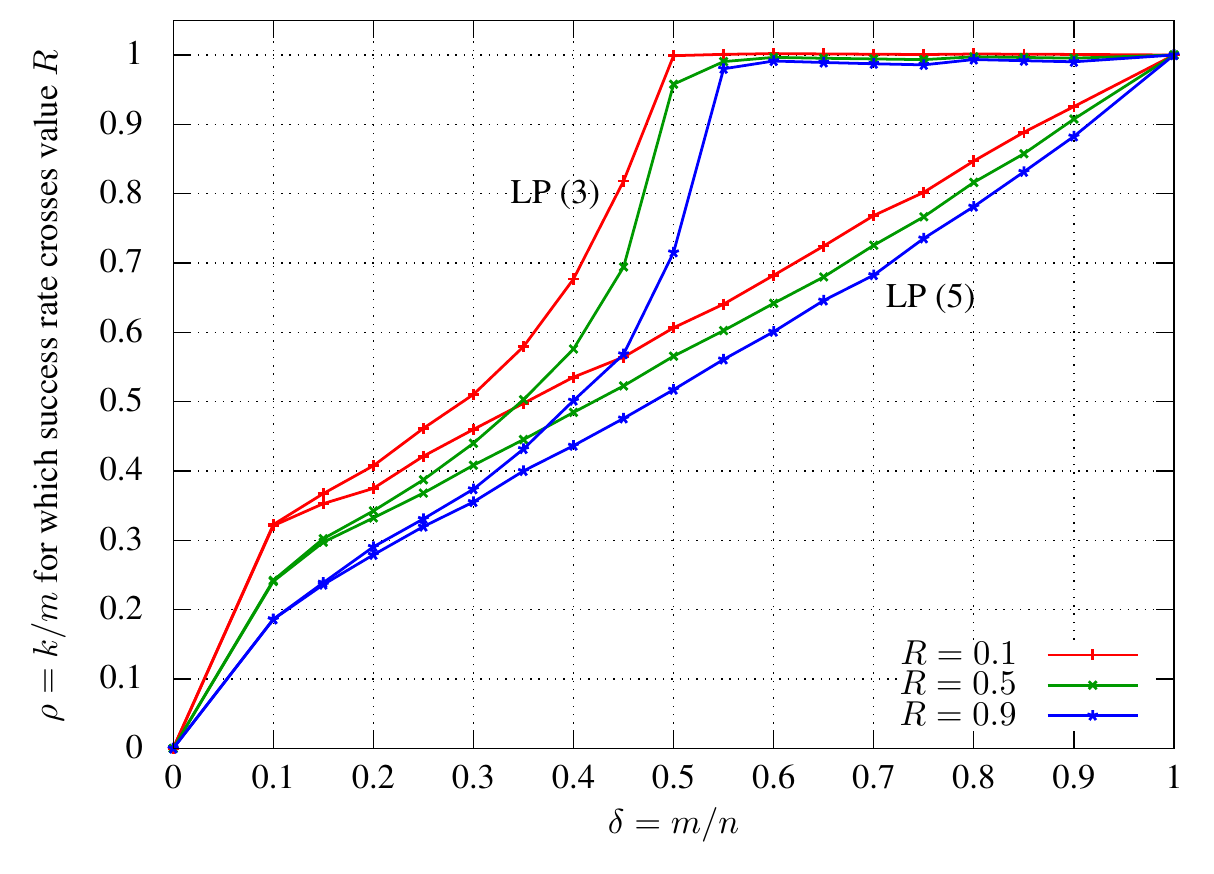}
  }                
  \subfloat[Transition behavior of LPs~\eqref{mlp},\eqref{lp} in $(\delta,\eta)$ space]{
    \label{fig:right}
    \includegraphics[width=0.49\textwidth]{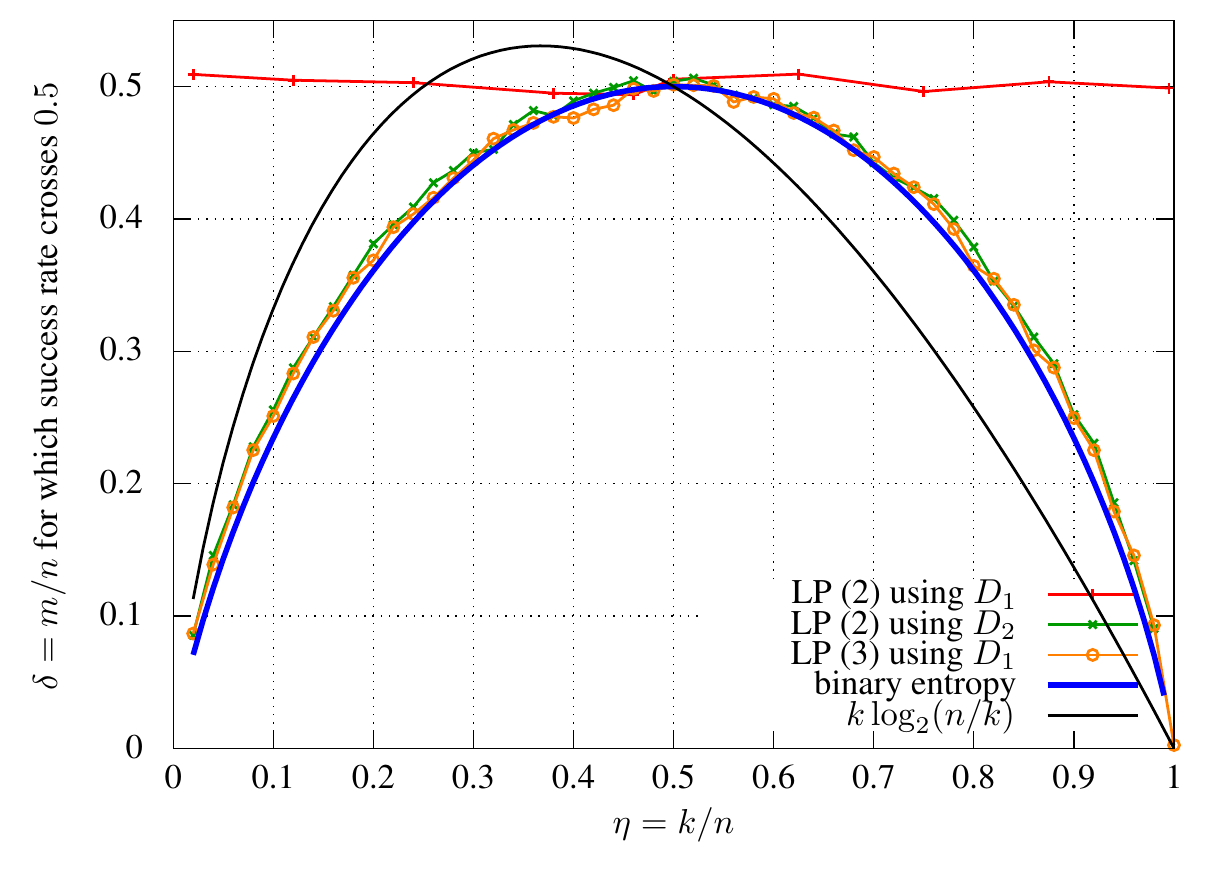}
  }                
  \vspace{-2mm}
  \caption{Simulation Results}
  \label{fig:res}
\end{figure}

For the first set of experiments, following~\cite{donoho2010precise}, we vary $m$ from $n/10$ to $9n/10$ in $17$ equal steps and for each $n, m$ we vary $k$ from $1$ to $m$ in $m/4$ equal steps. We consider the experimental data on a $3D$ space where the \emph{undersampling factor} $\delta = m/n$ is on the X-axis, the \emph{relative sparsity factor} $\rho = k/m$ on the Y-axis, and the success rates of LPs on the Z-axis. Fig.~\ref{fig:res}\subref{fig:left} shows the label sets for three different success rates $0.1, 0.5, 0.9$ for $n=500$ using distribution $D_1$. The lower three curves correspond to LP~\eqref{dlp1} and the upper three to LP~\eqref{lp}. For each LP, the three label sets represents a narrow \emph{transition zone} below which the success rate is almost 1 and above which success rate is almost 0. The transition zone becomes thinner as $n$ increases. We see that \eqref{lp} outperforms \eqref{dlp1}. 

For the second set of experiments, following~\cite{mangasarian2011probability}, we vary $k$ from $n/10$ to $9n/10$ in $17$ equal steps and vary $m$ from $0.02n$ to $0.98n$ in $25$ steps. Here again, we consider the experimental data on $3D$ space where \emph{absolute sparsity factor}, $\eta = k/n$ is on the X-axis, $\delta$ on the Y-axis and the success rates of LPs on the Z-axis. Fig.~\ref{fig:res}\subref{fig:right} shows the label sets for success rate $0.5$ for $n=500$ of LP~\eqref{mlp} using distributions $D_1, D_2$ and of LP~\eqref{lp} using $D_1$. Results for other distributions are similar and are not shown here. We call the value of $\delta$ for which success rate crosses $0.5$ the \emph{transition point}. If we consider the transition points as a function of $\eta$, then the label sets for LP~\eqref{lp} suggest the following conjecture:

\begin{conjecture}
  \label{cnj:binaryentropy} If the entries of $A$ are i.i.d. from some absolutely continuous distribution, then for large $n$, LP~\eqref{lp} exactly recovers a binary $k$-sparse solution to~\eqref{main} using $nH(k/n)/2$ measurements.  
\end{conjecture}

Fig.~\ref{fig:res}\subref{fig:right} also plots the function $k\log_2(n/k)$, which is a lower bound on the number of measurements needed to recover general $k$-sparse signals given by compressed sensing. We see that for $\eta <0.5$, LP~\eqref{lp} requires fewer measurements for binary signals.

\vspace{-5mm}

\vspace{-6mm}

{
\small
\bibliographystyle{unsrt}
\bibliography{mining}
}

%
%
%
%
%

\end{document}